\documentclass[aps,prd,onecolumn,groupedaddress,showpacs,nofootinbib,amssymb]{revtex4}
\usepackage[dvips]{graphicx}
\usepackage{amssymb}
\usepackage{amsmath}
\usepackage{graphicx,color}
\usepackage{amsfonts}
\usepackage{bm}
\usepackage{cancel}
\usepackage{comment}

\newcommand\be{\begin{equation}}
\newcommand\ee{\end{equation}}

\allowdisplaybreaks[4]

\begin{document}

\tolerance=5000

\title{Neutron Stars in Scalar-tensor Gravity with Quartic Order Scalar Potential}
\author{S.D.~Odintsov,$^{1,2}$\,\thanks{odintsov@ieec.uab.es}
V.K.~Oikonomou,$^{3,4}$\,\thanks{v.k.oikonomou1979@gmail.com}}
 \affiliation{$^{1)}$ ICREA, Passeig Luis Companys, 23, 08010 Barcelona, Spain\\
$^{2)}$ Institute of Space Sciences (IEEC-CSIC) C. Can Magrans
s/n,
08193 Barcelona, Spain\\
$^{3)}$ Department of Physics, Aristotle University of
Thessaloniki, Thessaloniki 54124,
Greece\\
$^{4)}$ Laboratory for Theoretical Cosmology, Tomsk State
University of Control Systems and Radioelectronics, 634050 Tomsk,
Russia (TUSUR)}

\tolerance=5000

\begin{abstract}
In this work we investigate the effects of a non-minimally coupled
quartic order scalar model on static neutron stars, with the
non-minimal coupling in the Jordan frame being of the form $f(\phi
) = 1 + \xi\phi^2$. Particularly we derive the Einstein frame
Tolman-Oppenheimer-Volkoff equations, and by numerically
integrating them for both the interior and the exterior of the
neutron star, using a double shooting python 3 based numerical
code, we extract the masses and radii of the neutron stars
evaluated finally in the Jordan frame, along with several other
related physical quantities of interest. With regard to the
equation of state for the neutron star, we use a piecewise
polytropic equation of state with the central part being
Skyrme-Lyon (SLy), Akmal-Pandharipande-Ravenhall (APR) or the
Wiringa-Fiks-Fabrocini (WFF1) equations of state. The resulting
$M-R$ graphs are compatible with the observational bounds imposed
by the GW170817 event which require the radius of a static $M\sim
1.6 M_{\odot}$ neutron star to be larger than
$R=10.68^{+15}_{-0.04}$km and the radius of a static neutron star
corresponding to the maximum mass of the star to be larger than
$R=9.6^{+0.14}_{-0.03}$km. Moreover, the WFF1 EoS, which was
excluded for static neutron stars in the context of general
relativity, for the a quartic order scalar model neutron star
model provides realistic results compatible with the GW170817
event.
\end{abstract}


\maketitle

\section*{Introduction}

Neutron stars (NS) have developed to be in the epicenter of
current scientific interest, since they are literally laboratories
in the cosmos, for many scientific disciplines like nuclear
\cite{Tolos:2020aln,Lattimer:2012nd,Steiner:2011ft,Horowitz:2005zb,Watanabe:2000rj,Shen:1998gq,Xu:2009vi,Hebeler:2013nza,Mendoza-Temis:2014mja,Ho:2014pta,Kanakis-Pegios:2020kzp}
and high energy particle physics
\cite{Buschmann:2019pfp,Safdi:2018oeu,Hook:2018iia,Edwards:2020afl,Nurmi:2021xds},
modified gravity
\cite{Astashenok:2020qds,Capozziello:2015yza,Astashenok:2014nua,Astashenok:2014pua,Astashenok:2013vza,Arapoglu:2010rz,Astashenok:2020cqq}
and astrophysics
\cite{Sedrakian:2015krq,Khadkikar:2021yrj,Sedrakian:2006zza,Sedrakian:2018kdm,Bauswein:2020kor,Vretinaris:2019spn,Bauswein:2020aag,Bauswein:2017vtn,Most:2018hfd,Rezzolla:2017aly}.
Nearly fifty years after the first observation of a NS by Jocelyn
Bell, the observational aspects of neutron stars have been
developed quite significantly, with the LIGO-Virgo collaboration
being the ``tip of the spear'' in observing and analyzing
gravitational waves emerging from NSs and black holes processes.
Thus the aim of understanding the inner processes of neutron stars
physics, which was scientifically a dream some decades ago, has
developed to be every day physics nowadays. The LIGO-Virgo
collaboration has set the stage for future discoveries, and the
upcoming LISA collaboration is expected to further improve our
knowledge on astrophysical and cosmological gravitational waves.
Even the early results of the LIGO-Virgo collaboration has proven
gold for both astrophysics and theoretical astrophysics, for
example the pioneering 2017 observation known as GW170817 event
\cite{TheLIGOScientific:2017qsa} has indicated that the
gravitational waves propagate with a speed that is nearly equal to
that of light's in vacuum. This result was obtained because
luckily the GW170817 event was accompanied by a kilonova
originating electromagnetic radiation which arrived almost
simultaneously with the gravitational waves. From  a theoretical
point of view, this observation has excluded quite many
theoretical cosmology models which predicted a gravitational wave
speed different than that of light's, although rectifications of
problematic theories can be constructed in order for them to
comply with the GW170817 event, see for example
\cite{Odintsov:2020sqy,Oikonomou:2020sij}. In addition, the
GW190814 event \cite{Abbott:2020khf} had a mysterious secondary
component with mass in the range of the so-called mass-gap region.
If this secondary component is proved to be a neutron star, or
even a black hole, it will be a result of fundamental importance.
In addition, fundamental particle physics models could be tested
from pulsars observations, like the axion models that may be
transformed to photons in the strong atmospheric magnetic fields
of pulsars \cite{Hook:2018iia}. Of course, observations may
provide useful insights towards the understanding of nuclear
matter equation of state (EoS) at high compression and at
supranuclear densities. For a mainstream of textbooks and reviews
on neutron stars, we refer the reader to Refs.
\cite{Haensel:2007yy,Friedman:2013xza,Baym:2017whm,Lattimer:2004pg,Olmo:2019flu}.

In view of the GW190814 event, the possibility of having a neutron
star in the mass-gap region is rather scientifically stimulating,
since such a result is marginally supported by general relativity
(GR) even for the stiffest equation of state. For some interesting
perspectives for the GW190814 event see
\cite{Nathanail:2021tay,Koppel:2019pys}. In the literature there
exist several EoSs that may predict maximum masses for NSs that
could describe the GW190814 event, but only marginally for the
moment. However, if future observations indicate the existence of
NSs masses in the mass-gap region, above $M\geq 2.6 M_{\odot}$,
this will indicate that GR may be complemented theoretically by an
alternative description. Modified gravity
\cite{Nojiri:2017ncd,Capozziello:2011et,Capozziello:2010zz,Nojiri:2006ri,
Nojiri:2010wj,delaCruzDombriz:2012xy,Olmo:2011uz,dimo} in its
various forms may play some fundamental role if this scenario
proves true, and in the literature there exist several
descriptions of NSs with masses that cannot be reached by even the
stiffest GR EoSs \cite{Astashenok:2014nua}. In addition, in the
context of modified gravity, some puzzles of ordinary GR
descriptions of NSs, may be consistently resolved, such as the
hyperon puzzle \cite{Astashenok:2014pua}. The inner core of
neutron stars is a mystery, and the existence of hyperons is
softening significantly the EoS (if the two hyperon interaction is
only taken into account), thus a large mass neutron star will
indicate that a hyperon-free EoS must be used, but modified
gravity can also harbor hyperon-related EoSs and simultaneously
producing large masses for NSs \cite{Astashenok:2014pua}. It is
interesting to note though that for static neutron stars, even in
the context of modified gravity one does not expect the maximum
masses to be larger than $3\,M_{\odot}$ \cite{Astashenok:2021peo}.

Modified gravity can serve as a successful complement of GR when
large mass NSs are considered. In cosmology the urge for a
modified gravity description of several phenomena is compelling,
since dark energy and some aspects of inflation cannot be
described by GR. With regard to dark energy, the most successful
description of it, the $\Lambda$-Cold-Dark-Matter model heavily
relies on the existence of a cosmological constant, which is
constant so the dark energy EoS is exactly on the phantom divide
line. However, the latest Planck data on cosmological parameters
\cite{Aghanim:2018eyx} indicate that range of values that the dark
energy EoS, namely $\omega_{DE}$, is allowed to take, crosses the
phantom divide line marginally towards the phantom regime, and
specifically $\omega_{DE}=-1.018\pm 0.031$. Such a phantom
possibility can of course be described by GR, but only by using a
phantom scalar field \cite{Caldwell:2003vq}, which is not an
elegant and self-consistent description for a physical system.
With regard to inflation, it is basically a classical era, a
post-quantum era however, very close chronically and energy-wise
to the quantum epoch of our Universe. The most popular description
of inflation makes use of a scalar field, which is basically a GR
description, but it is highly likely that the quantum era might
leave its imprints on the inflationary Lagrangian, in terms of
non-minimal couplings of scalars with gravity, couplings of
scalars with higher order curvature terms, or Gauss-Bonnet
couplings or even higher order curvature terms, see the reviews
\cite{Nojiri:2017ncd,Capozziello:2011et,Capozziello:2010zz,Nojiri:2006ri,
Nojiri:2010wj,delaCruzDombriz:2012xy,Olmo:2011uz,dimo} for a
complete description of all the possibilities for the effective
inflationary Lagrangian.

In this paper we shall consider one of the above aspects for the
Lagrangian of gravity, that of non-minimal couplings of scalar
fields on gravity. We shall thus investigate scalar-tensor
theories with non-minimal coupling of the scalar field with the
Ricci scalar. We aim to study the hydrodynamic stability of static
NSs for a specific physically motivated Lagrangian, and to find
the quantities related to the hydrodynamic stability of NSs. The
scalar-tensor theory we shall use is related to a quartic order
scalar potential, which as a scalar-tensor theory is quite popular
due to its similarity to the Higgs inflationary potential in
cosmological contexts
\cite{Bezrukov:2014bra,GarciaBellido:2011de,Bezrukov:2010jz,Bezrukov:2007ep,Mishra:2018dtg,Steinwachs:2013tr,Rubio:2018ogq,Kaiser:1994vs,Gundhi:2018wyz,CervantesCota:1995tz,Kamada:2012se,Schlogel:2014jea,Fuzfa:2013yba},
since it can generate a successful inflationary era compatible
with the latest Planck data on inflation \cite{Akrami:2018odb}. As
a scalar theory, our motivation for studying the a quartic order
scalar model is not accidental at all. Thus, since its
inflationary aspects provide a successful description of
inflation, in this paper we shall consider the implications of the
quartic order scalar model Lagrangian on static NSs. Our approach
and notation will be technically similar to many studies of
theoretical astrophysics approaches for scalar tensor theories
\cite{Pani:2014jra,Staykov:2014mwa,Horbatsch:2015bua,Silva:2014fca,Doneva:2013qva,Xu:2020vbs,Salgado:1998sg,Shibata:2013pra,Arapoglu:2019mun,Ramazanoglu:2016kul,AltahaMotahar:2019ekm,Chew:2019lsa,Blazquez-Salcedo:2020ibb,Motahar:2017blm},
and we shall use the notation of such approaches in order to
comply with the existing literature. For a recent similar work in
the same spirit as the present article see
\cite{Odintsov:2021qbq}. So we shall extract the
Tolman-Oppenheimer-Volkoff (TOV) equations in the Einstein frame,
and we shall calculate the gravitational masses of the NSs and the
radii for several piecewise polytropic equations of state which
are valid up to large central densities
\cite{Read:2008iy,Read:2009yp}. Our results will be the Jordan
frame quantities, regarding the circumferential radii of the NSs,
and with regard to the mass, since there are several different
possibilities for the definition of the gravitational mass, such
as the Komar mass \cite{Komar:1958wp}, in this paper we shall use
the Arnowitt-Deser-Misner (ADM) mass \cite{Arnowitt:1960zzc},
which for static stellar configurations coincide
\cite{Shibata:2013ssa}. For the solution of the TOV equations, we
shall use the well known numerical code pyTOV-STT \cite{niksterg},
appropriately modified to incorporate the scalar potential, and we
shall use a double shooting method in order to most accurate
values for the scalar field and one of the two metric related
functions. In addition, before getting to the core of this study,
we shall present in brief the cosmologist's and theoretical
astrophysicist's perspective of scalar-tensor theories, since
there is a difference in notation, which might cause confusion
initially to a cosmologist, but for the calculations we shall use
the usual theoretical astrophysics notation and physical units,
with all the quantities of interest converted in the end in the
CGS system. As for the values of the free parameters, these shall
be aligned with the values imposed by the inflationary
constraints. We need to mention that NSs with similar potential in
the Jordan frame were studied in \cite{Arapoglu:2019mun}, but our
approach is different since we deal with the Einstein frame
theory, plus we use a physically motivated from inflation Einstein
frame scalar field potential.

This paper is organized as follows: In section II we present the
cosmologist's perspective of scalar-tensor theory in the Jordan
and Einstein frame, where the natural units physical system is
used. In section II, we present the theoretical astrophysicist's
perspective of the scalar-tensor gravity in geometrized units, and
we shall try to bridge the two approaches in order to make direct
correspondence with the inflationary quartic order scalar model,
always in geometrized units. In section III, we describe in brief
the essential features of the piecewise polytropic equation of
state approach, and in section IV we study in detail the quartic
order scalar model potential in the Einstein frame for compact
static spherically symmetric stellar objects. We derive the TOV
equations in the Einstein frame and by using a well-known
python-based numerical code, which we appropriately modified to
incorporate the scalar potential, we numerically solve the TOV
equations using a double shooting method for maximum accuracy of
the delivered results. In the same section we present the results
of our analysis for several piecewise polytropic EoSs of interest,
and for various values of the free parameters, and finally we
directly compare the results with the GR ones. Also we examine the
values of the scalar field in the Jordan frame and we investigate
whether the approximations needed are satisfied by the numerical
results. Finally, the conclusions follow at the end of the paper.

\section{Scalar-tensor Gravity from the Cosmologist's Perspective}

The notation used for scalar-tensor theory in cosmological
contexts is different from the one used in theoretical
astrophysics context. Therefore, we shall present both frameworks
in order to bridge the gap between the two contexts, since we
shall make a direct correspondence of the quartic order scalar
inflationary model from the cosmological context to a theoretical
astrophysics context. In this section we shall present the
cosmologist's perspective of scalar-tensor theories, using natural
units. Details for the formalism used can be found in Refs.
\cite{Kaiser:1994vs,valerio,Faraoni:2013igs,Buck:2010sv,Faraoni:1998qx,Odintsov:2021qbq}.

Consider the Jordan frame action in the presence of the
non-minimally coupled inflaton and the matter Lagrangian,
\begin{equation}\label{c1}
\mathcal{S}_J=\int d^4x\Big{[}f(\phi)R-\frac{1}{2}g^{\mu
\nu}\partial_{\mu}\phi\partial_{\nu}\phi-U(\phi)\Big{]}+S_m(g_{\mu
\nu},\psi_m)\, ,
\end{equation}
where $\psi_m$ denotes the perfect matter fluids present in the
Jordan frame, with pressure $P$ and energy density $\epsilon$, and
$g_{\mu \nu}$ the Jordan frame metric. The minimal coupling choice
corresponds to,
\begin{equation}\label{c2}
f(\phi)=\frac{1}{16 \pi G}=\frac{M_p^2}{2}\, ,
\end{equation}
where,
\begin{equation}\label{c3}
M_p=\frac{1}{\sqrt{8\pi G}}\, ,
\end{equation}
is the Jordan frame reduced Planck mass, which is $M_p=2.43\times
10^{18}$GeV, and $G$ is Newton's gravitational constant in the
Jordan frame. Now we shall perform a conformal transformation of
the following form,
\begin{equation}\label{c4}
\tilde{g}_{\mu \nu}=\Omega^2g_{\mu \nu}\, ,
\end{equation}
or equivalently,
\begin{equation}\label{c5}
\tilde{g}^{\mu \nu}=\Omega^{-2}g^{\mu \nu}
\end{equation}
in order to obtain the action in the Einstein frame. Hereafter,
the tilde will denote quantities in the Einstein frame. Also
$\Omega$ in terms of $f(\phi)$ is written as
\cite{Kaiser:1994vs,Mishra:2018dtg,valerio},
\begin{equation}\label{c6}
\Omega^2=\frac{2}{M_p^2}f(\phi)\, .
\end{equation}
The quantities appearing in the Jordan frame action are
transformed in the following way,
\begin{equation}\label{c7}
\sqrt{-g}=\Omega^{-4}\sqrt{-\tilde{g}}\, ,
\end{equation}
the Ricci scalar transforms as follows,
\begin{equation}\label{c8}
R=\Omega^2\left(\tilde{R}+6\tilde{\square}f-6\tilde{g}^{\mu
\nu}f_{\mu}f_{\nu}\right)\, ,
\end{equation}
and the d'Alembertian is,
\begin{equation}\label{c9}
\tilde{\square}f=\frac{1}{\sqrt{-\tilde{g}}}\tilde{\partial}_{\mu}\left(
\sqrt{-\tilde{g}}\tilde{g}^{\mu \nu}\partial_{\nu}f\right)\, .
\end{equation}
Thus the non-minimal coupling term is transformed in the Einstein
frame as follows,
\begin{equation}\label{c10}
\int d^4x\sqrt{-g}f(\phi)R\to \int d^4x
\sqrt{-\tilde{g}}\frac{M_p^2}{2}\left(
\tilde{R}-6\left(\frac{1}{\Omega^2}\right)^2\tilde{g}^{\mu
\nu}\tilde{\partial}_{\mu}\Omega \tilde{\partial}_{\nu}\Omega
\right)\, ,
\end{equation}
and the Jordan frame kinetic term plus the potential term
transform as follows,
\begin{equation}\label{c11}
\int d^4 x \sqrt{-g} \Big{[}-\frac{1}{2}g^{\mu
\nu}\partial_{\mu}\phi \partial{\nu}-U(\phi)\Big{]} +S_m(g_{\mu
\nu},\psi_m)\to \int d^4 x
\sqrt{-\tilde{g}}\Big{[}-\frac{1}{2\Omega^2} \tilde{g}^{\mu
\nu}\tilde{\partial}_{\mu}\phi
\tilde{\partial}{\nu}\phi-\frac{U(\phi)}{\Omega^4(\varphi)}\Big{]}
+S_m(\Omega^2\tilde{g}_{\mu \nu},\psi_m)
\end{equation}
thus in the end, the Einstein frame action in terms of
$\Omega^2=\frac{2}{M_p^2}f$  becomes,
\begin{equation}\label{c12}
\mathcal{S}_E=\int
d^4x\sqrt{-\tilde{g}}\Big{[}\frac{M_p^2}{2}\tilde{R}-\frac{\zeta
(\phi)}{2} \tilde{g}^{\mu \nu }\tilde{\partial}_{\mu}\phi
\tilde{\partial}_{\nu}\phi-V(\phi)\Big{]}+S_m(\Omega^{-2}\tilde{g}_{\mu
\nu},\psi_m)\, ,
\end{equation}
where,
\begin{equation}\label{c13}
V(\phi)=\frac{U(\phi)}{\Omega^4}\, ,
\end{equation}
and also we introduced $\zeta(\phi)$ which is,
\begin{equation}\label{c14}
\zeta
(\phi)=\frac{3}{2}M_p^2\frac{1}{f^2}\Big{(}\frac{df}{d\phi}\Big{)}^2+\frac{M_p^2}{2f}\,
.
\end{equation}
In addition, clearly the Einstein frame scalar field $\phi$ is not
canonical since $\zeta (\phi)\neq 1$, hence we perform the
following rescaling,
\begin{equation}\label{c15}
\Big{(}\frac{d\varphi}{d \phi}\Big{)}^2 =\zeta(\phi)\, ,
\end{equation}
or equivalently,
\begin{equation}\label{c16}
\frac{d \varphi}{d \phi}=
M_p\sqrt{\frac{1}{2f}+\frac{3}{2}\Big{(}\frac{f'}{f}\Big{)}^2}\, ,
\end{equation}
where the prime indicates differentiation with respect to $\phi$,
that is $f'=\frac{df}{d\phi}$. Hence the Einstein frame action in
terms of the canonical scalar field $\varphi$ reads,
\begin{equation}\label{c17}
\mathcal{S}_E=\int
d^4x\sqrt{-\tilde{g}}\Big{[}\frac{M_p^2}{2}\tilde{R}-\frac{1}{2}\tilde{g}^{\mu
\nu } \tilde{\partial}_{\mu}\varphi
\tilde{\partial}_{\nu}\varphi-V(\varphi)\Big{]}+S_m(\Omega^2\tilde{g}_{\mu
\nu},\psi_m)
\end{equation}
where,
\begin{equation}\label{c18}
V(\varphi)=\frac{U(\varphi)}{\Omega^4}=\frac{U(\varphi)}{4
M_p^4f^2}\, .
\end{equation}
Clearly, the fluids $\psi_m$ in the Einstein frame are not
perfect, due to the presence of the conformal factor in the matter
action, hence, the energy momentum tensor which is,
\begin{equation}\label{c19}
\tilde{T}_{\mu \nu}=-\frac{2}{\sqrt{-\tilde{g}}}\frac{\delta
L_m}{\delta \tilde{g}^{\mu \nu}}\, ,
\end{equation}
from the Jordan frame to the Einstein frame transforms as,
\begin{equation}\label{c20}
\tilde{T}_{\mu \nu}=\Omega^{-2}(\varphi)T_{\mu \nu}\, ,
\end{equation}
\begin{equation}\label{c21}
\tilde{T}^{\mu}_{\nu}=\Omega^{-4}(\varphi)T^{\mu}_{\nu}\, ,
\end{equation}
\begin{equation}\label{c22}
\tilde{T}^{\mu \nu}=\Omega^{-6}(\varphi)T^{\mu \nu}\, ,
\end{equation}
and the trace of the energy momentum tensor transforms as,
\begin{equation}\label{c23}
\tilde{T}=\Omega^{-4} T \, .
\end{equation}
The continuity equation for the energy momentum tensor in the
Einstein frame reads,
\begin{equation}\label{c24}
\tilde{\partial}^{\mu}\tilde{T}_{\mu \nu}=-\frac{d}{d\varphi}[\ln
\Omega]\tilde{T}\tilde{\partial}_{\nu}\phi\, .
\end{equation}
In the Jordan frame, where the matter fluids are perfect fluids,
the energy momentum tensor takes the form,
\begin{equation}\label{c25}
T_{\mu \nu}=(P+\varepsilon)u_{\mu}u_{\nu}+Pg_{\mu \nu}\, ,
\end{equation}
where $P$ and $\varepsilon$ are the Jordan frame pressure and
energy momentum respectively, and suppose that in the Einstein
frame the energy momentum tensor is,
\begin{equation}\label{c26}
\tilde{T}_{\mu
\nu}=(\tilde{P}+\tilde{\varepsilon})\tilde{u}_{\mu}\tilde{u}_{\nu}+\tilde{P}\tilde{g}_{\mu
\nu}\, ,
\end{equation}
where the Einstein frame pressure is denoted as $\tilde{P}$ and
the corresponding Einstein frame energy density is
$\tilde{\varepsilon}$. The four-velocity $\tilde{u}_{\mu}$
satisfies,
\begin{equation}\label{c27}
\tilde{g}_{\mu \nu}\tilde{u}^{\mu}\tilde{u}^{\nu}=-1\, .
\end{equation}
Thus, the four-velocity transforms as
$\tilde{u}_{\mu}=\Omega^{1}u_{\mu}$, and by direct comparison of
the energy momentum tensors in the Jordan and the Einstein frame,
we get that the pressure and the energy density in the two frames
are related as follows,
\begin{equation}\label{c28}
\tilde{\varepsilon}=\Omega^{-4}(\varphi)\varepsilon,\,\,\,\tilde{P}=\Omega^{-4}(\varphi)P\,
.
\end{equation}

\section{Scalar-tensor Gravity from the Theoretical Astrophysicist's Perspective}

In the context of theoretical astrophysics, a different notation
for the gravitational action and conformal transformation is used,
leading to different dimensions of the scalar field, since the
Geometrized units are used usually ($G=c=1$). This formalism and
notation is commonly used in most of the theoretical astrophysics
works. In the following we shall use the notation and formalism of
\cite{Pani:2014jra}, with the only difference being that the tilde
notation will indicate quantities in the Einstein frame, contrary
to Ref. \cite{Pani:2014jra}, where the Jordan frame quantities are
denoted with a ``tilde''.

To start with consider the non-minimally coupled scalar field
action in the Jordan frame (note our difference in the usage of
the tilde, in order to provide a uniform notation for the present
work) \cite{Pani:2014jra},
\begin{equation}\label{ta}
\mathcal{S}=\int d^4x\frac{\sqrt{-g}}{16\pi
G}\Big{[}f(\phi)R-\frac{1}{2}g^{\mu
\nu}\partial_{\mu}\phi\partial_{\nu}\phi-U(\phi)\Big{]}+S_m(\psi_m,g_{\mu
\nu})\, ,
\end{equation}
where $\phi$ denotes the scalar field in the Jordan frame. We
shall use Geometrized units in which $c=G=1$, and we shall make
the conformal transformation,
\begin{equation}\label{ta1}
\tilde{g}_{\mu \nu}=A^{-2}g_{\mu \nu}\, ,
\end{equation}
where recall that the tilde denotes quantities in the Einstein
frame. The function $A(\phi)$ is arbitrary, but for the choice,
\begin{equation}\label{ta2}
A(\phi)=f^{-1/2}(\phi)\, ,
\end{equation}
one obtains a minimally coupled scalar field in the Einstein
frame. So from now on we shall assume that
$A(\phi)=f^{-1/2}(\phi)$, and in addition, the scalar potential in
the Einstein frame is,
\begin{equation}\label{ta3}
V(\phi)=\frac{U(\phi)}{f^2(\phi)}\, ,
\end{equation}
but the resulting action contains a non-canonical kinetic term for
the scalar field $\phi$, which can be made canonical by making the
following simple transformation,
\begin{equation}\label{ta4}
\Big{(}\frac{d \varphi }{d \phi}\Big{)}=\frac{1}{\sqrt{4\pi}}
\sqrt{\Big{(}\frac{3}{4}\frac{1}{f^2}\Big{(}\frac{d
f}{d\phi}\Big{)}^2+\frac{1}{4f}\Big{)}}\, ,
\end{equation}
where $\varphi$ is the canonical scalar field in the Einstein
frame, hence the gravitational action in the Einstein frame
becomes in terms of the canonical scalar field $\varphi$,
\begin{equation}\label{ta5}
\mathcal{S}=\int
d^4x\sqrt{-\tilde{g}}\Big{(}\frac{\tilde{R}}{16\pi}-\frac{1}{2}
\tilde{g}_{\mu \nu}\partial^{\mu}\varphi
\partial^{\nu}\varphi-\frac{V(\varphi)}{16\pi}\Big{)}+S_m(\psi_m,A^2(\varphi)g_{\mu
\nu})\, .
\end{equation}
In the Einstein frame, the field equations for a general metric
$\tilde{g}_{\mu \nu}$ read,
\begin{equation}\label{ta6}
\tilde{G}_{\mu \nu}=8\pi \tilde{T}_{\mu \nu}+8 \pi
\Big{(}\tilde{\partial}_{\mu}\varphi
\tilde{\partial}_{\nu}\varphi-\frac{\tilde{g}_{\mu
\nu}}{2}\tilde{\partial}_{\sigma}\varphi
\tilde{\partial}^{\sigma}\varphi \Big{)}-\frac{\tilde{g}_{\mu
\nu}}{2}V(\varphi)\, ,
\end{equation}
\begin{equation}\label{ta7}
\tilde{\square}\varphi=-\frac{A'(\varphi)}{A(\varphi)}\tilde{T}+\frac{V'(\varphi)}{16\pi}\,
,
\end{equation}
where $\tilde{T}$ is the trace of the energy momentum tensor in
the Einstein frame. The energy momentum tensor in the Einstein
frame $\tilde{T}^{\mu \nu}$ is related to the perfect fluid energy
momentum tensor of the Jordan frame $T_{\mu \nu}$ as follows,
\begin{equation}\label{ta8}
\tilde{T}_{\mu \nu}=A^2T_{\mu \nu}\, ,
\end{equation}
\begin{equation}\label{ta9}
\tilde{T}^{\mu}_{\nu}=A^4T^{\mu}_{\nu}\, ,
\end{equation}
and the traces are related as follows,
\begin{equation}\label{ta10}
\tilde{T}=A^4T\, ,
\end{equation}
therefore the pressure $\tilde{P}$ and the energy density
$\tilde{\epsilon}$ in the Einstein frame are related to the Jordan
frame ones $P$ and $\epsilon$ as follows,
\begin{equation}\label{ta11}
\tilde{\epsilon}=A^4\epsilon \, ,
\end{equation}
\begin{equation}\label{ta12}
\tilde{P}=A^4 P \, .
\end{equation}
In the cosmological applications, in natural units ($c=\hbar=1$),
the various physical quantities have the following dimensions,
\begin{align}\label{ta13}
&x^{\mu}=[m]^{-1}\, ,\,\,\, \tilde{\partial}_{\mu}=[m]\\
\notag & \tilde{R}=[m]^{2}\, ,\,\,\, \frac{1}{G}=[m]^{2}\\ \notag
& V(\varphi)=[m]^{4}\, , \,\,\,\varphi=[m]
\end{align}
In the astrophysical contexts, where the more appropriate
Geometrized Units are used ($G=c=1$), the scalar field is
dimensionless. The TOV equations eventually will be presented in
the Geometrized Units. If we start from the action of the
cosmological contexts, namely,
\begin{equation}\label{ta14}
\mathcal{S}_E=\int
d^4x\sqrt{-\tilde{g}}\Big{[}\frac{M_p^2}{2}\tilde{R}-\frac{1}{2}\tilde{g}^{\mu
\nu } \tilde{\partial}_{\mu}\varphi
\tilde{\partial}_{\nu}\varphi-V(\varphi)\Big{]}\, ,
\end{equation}
it can be brought in the form,
\begin{equation}\label{ta15}
\mathcal{S}_E=\int d^4x\sqrt{-\tilde{g}}\frac{1}{16\pi G}
\Big{[}\tilde{R}-\frac{1}{2}\tilde{g}^{\mu \nu
}\tilde{\partial}_{\mu}\tilde{\varphi}
\tilde{\partial}_{\nu}\tilde{\varphi}-V(\tilde{\varphi})\Big{]}\,
,
\end{equation}
by rescaling the scalar field $\varphi$ as follows,
$$\tilde{\varphi}=\sqrt{16\pi G}\varphi=\frac{\sqrt{2}}{M_p} \varphi \, .$$
The above notation for the action is unusual, meaning that the
factor $1/(16\pi G)$ does not multiply the whole action, just the
Ricci scalar. In theoretical astrophysics contexts, the action
appears with an overall multiplicative factor $1/(16\pi G)$, so
this is what we also adopt for this work.

\section{The Piecewise Polytropic Equation of State}

For the present NS study in Einstein frame we shall consider a
piecewise polytropic EoS \cite{Read:2008iy,Read:2009yp} (see also
the introductory text of the TOV-pp code which can be found in
\cite{niksterg}), which we now describe in brief. A piecewise
polytropic EoS consists of a low-density part $\rho<\rho_0$, which
in general can be a tabulated crust EoS, and of a high density
part with $\rho\gg \rho_0$. The density $\rho_0$ is a matching
density between the low and high density pieces, and the piecewise
polytropic EoS needs other two dividing high densities, $\rho_1 =
10^{14.7}{\rm g/cm^3}$ and  $\rho_2= 10^{15.0}{\rm g/cm^3}$. The
density and pressure in each of the piecewise density interval
$\rho_{i-1} \leq \rho \leq \rho_i$ satisfy the following the
polytropic relation,
\begin{equation}\label{pp1}
P = K_i\rho^{\Gamma_i}\, ,
\end{equation}
and the requirement is that continuity is necessary at the
crossing points of each piece. Particularly, at a dividing density
$\rho_i$, the following continuity relations must hold true,
\begin{equation}\label{pp2}
P(\rho_i) = K_i\rho^{\Gamma_i} = K_{i+1}\rho^{\Gamma_{i+1}}\, ,
\end{equation}
and from the above continuity relations we can obtain the
parameters $K_2$ and $K_3$ for a given chosen $K_1, \Gamma_1,
\Gamma_2, \Gamma_3$, or equivalently for a given initial pressure
$p_1$ and for given parameters $\Gamma_2$, and $\Gamma_3$, which
are not chosen arbitrarily. For the purposes article, the initial
pressure $p_1$ and the parameters $\Gamma_2$, and $\Gamma_3$ will
correspond to the values of three distinct EoSs, the WFF1
\cite{Wiringa:1988tp} which is a variational method EoS, the SLy
\cite{Douchin:2001sv} which is a potential method EoS, and the APR
EoS \cite{Akmal:1998cf}. Upon integrating the first law of
thermodynamics for barotropic fluids,
\begin{equation}\label{pp3}
d\frac{\epsilon}{\rho} = - P d\frac{1}{\rho}\, ,
\end{equation}
and the continuity requirement in the energy density, gives,
\begin{equation}\label{pp4}
\epsilon(\rho) = (1+\alpha_i)\rho +
\frac{K_i}{\Gamma_i-1}\rho^{\Gamma_i}\, ,
\end{equation}
for $\Gamma_i \neq 1$, where,
\begin{equation}\label{pp5}
\alpha_i = \frac{\epsilon(\rho_{i-1})}{\rho_{i-1}} -1 -
\frac{K_i}{\Gamma_i-1}\rho_{i-1}^{\Gamma_i-1}\, .
\end{equation}
Moreover, the sound speed $v_s =\sqrt{dP/d\epsilon}$ can be
expressed in terms of the parameters of the piecewise EoS,
\begin{equation}\label{pp6}
v_s(\rho) = \sqrt{\frac{\Gamma_i P}{\epsilon+P}}\, .
\end{equation}

\section{Neutron Stars in Scalar-tensor Gravity with Quartic Order Scalar Potential in the Einstein Frame}

Let us now proceed to the core of this study, and we shall present
the quartic order scalar model in the Jordan frame and the
corresponding Einstein frame theory, and accordingly we shall
present the field equations for the Einstein frame theory for a
spherically symmetric and static spacetime. We shall use the
astrophysical conventions and notation presented in section III,
but also we shall make contact with the cosmological notation in
order to make the appropriate choices for the values of the free
parameters. In the following we shall adopt the notation of
\cite{Pani:2014jra}. The quartic order scalar model as is used in
cosmological contexts \cite{Mishra:2018dtg}, has the following
Jordan frame action in Geometrized units ($G=1$ and we use the
notation of \cite{Pani:2014jra}),
\begin{equation}\label{ta}
\mathcal{S}=\int
d^4x\frac{\sqrt{-g}}{16\pi}\Big{[}f(\phi)R-\frac{1}{2}g^{\mu
\nu}\partial_{\mu}\phi\partial_{\nu}\phi-U(\phi)\Big{]}+S_m(\psi_m,g_{\mu
\nu})\, ,
\end{equation}
with the non-minimal coupling function $f(\phi)$ and the potential
$U(\phi)$ being defined as follows,
\begin{equation}\label{fofphi}
f(\phi)=1+\xi \phi^2\, ,
\end{equation}
\begin{equation}\label{jordanframepothiggs}
U(\phi)=\lambda \phi^4\, ,
\end{equation}
and with $\phi$ we obviously denote the Jordan frame scalar field.
Note that such choice of potentials corresponds to
multiplicatively-renormalizable scalar theory in curved spacetime
\cite{serg2}. Upon performing the conformal transformation,
\begin{equation}\label{ta1higgs}
\tilde{g}_{\mu \nu}=A^{-2}g_{\mu \nu}\, ,
\end{equation}
we obtain the Einstein frame action expressed in terms of the
canonical scalar field $\varphi$,
\begin{equation}\label{ta5higgs}
\mathcal{S}=\int
d^4x\sqrt{-\tilde{g}}\Big{(}\frac{\tilde{R}}{16\pi}-\frac{1}{2}
\tilde{g}_{\mu \nu}\partial^{\mu}\varphi
\partial^{\nu}\varphi-\frac{V(\varphi)}{16\pi}\Big{)}+S_m(\psi_m,A^2(\varphi)g_{\mu
\nu})\, ,
\end{equation}
and recall the ``tilde'' denotes Einstein frame quantities. For
the quartic order scalar model, the function $A(\phi)$ appearing
in the conformal transformation (\ref{ta1higgs}), is defined as
follows,
\begin{equation}\label{ta2higgs}
A(\phi)=f^{-1/2}(\phi)\, ,
\end{equation}
thus by using Eq. (\ref{fofphi}) we have,
\begin{equation}\label{ta2higgsini}
A(\phi)=\left(1+\xi \phi^2 \right)^{-1/2}\, ,
\end{equation}
and in the end we shall express the function $A(\phi)$ as a
function of the Einstein frame canonical scalar field $\varphi$,
after we obtain the relation $\phi (\varphi)$. Accordingly, the
Einstein frame potential is,
\begin{equation}\label{ta3higgs}
V(\phi)=\frac{U(\phi)}{f^2(\phi)}\, ,
\end{equation}
which in term of $\phi$ is,
\begin{equation}\label{potinitial}
V(\phi)=\frac{\lambda\phi^4}{\left(1+\xi \phi^2 \right)^2}\, ,
\end{equation}
and in the end we shall express the potential as a function of the
Einstein frame canonical scalar field $\varphi$. The relation
between the Jordan frame scalar field $\phi$ and the Einstein
frame canonical scalar field $\varphi$ is,
\begin{equation}\label{ta4higgs}
\frac{d \varphi }{d \phi}=\frac{1}{\sqrt{4\pi}}
\sqrt{\Big{(}\frac{3}{4}\frac{1}{f^2}\Big{(}\frac{d
f}{d\phi}\Big{)}^2+\frac{1}{4f}\Big{)}}\, ,
\end{equation}
so by substituting $f(\phi)$ from Eq. (\ref{fofphi}), we get,
\begin{equation}\label{finalrelationdiffphivaphi}
\frac{d \varphi }{d
\phi}=\frac{1}{\sqrt{16\pi}}\frac{\sqrt{1+\xi\phi^2+12\xi^2\phi^2}}{1+\xi\phi^2}\,
.
\end{equation}
Standard approximations used in cosmological contexts for the
above Eq. (\ref{finalrelationdiffphivaphi}) are the following (see
Eq. (\ref{inflconstraint}) later on in this section, and the
relevant discussion below it),
\begin{equation}\label{approxmain1}
\xi^2\phi^2\gg 1\, ,
\end{equation}
and simultaneously,
\begin{equation}\label{approxmain2}
\xi^2\phi^2\gg \xi \phi^2\, ,
\end{equation}
with Eq. (\ref{approxmain2}) holding automatically true for
$\xi\gg 1$. Using the approximations (\ref{approxmain1}) and
(\ref{approxmain2}), Eq. (\ref{finalrelationdiffphivaphi})
becomes,
\begin{equation}\label{final1}
\frac{d \varphi }{d \phi}\simeq
\frac{\sqrt{12}}{\sqrt{16\pi}}\frac{\xi\phi}{1+\xi\phi^2}=\frac{\sqrt{12}}{2\sqrt{16\pi}}\frac{f'(\phi)}{f(\phi)}\,
,
\end{equation}
thus upon integration, Eq. (\ref{final1}) yields the relation
between $\varphi$ and $\phi$ which is,
\begin{equation}\label{finalvarphiphirelation}
\varphi=\frac{\sqrt{12}}{2\sqrt{16\pi}}\ln \left( f(\phi)
\right)=\frac{\sqrt{12}}{2\sqrt{16\pi}}\ln \left( 1+\xi
\phi^2\right)\, ,
\end{equation}
hence,
\begin{equation}\label{oneplusxiphisquare}
1+\xi\phi^2=e^{\frac{2\sqrt{16\pi}}{\sqrt{12}}\, \varphi}\, .
\end{equation}
Thus from Eq. (\ref{ta2higgsini}), we can express $A(\phi)$ as a
function of the Einstein frame canonical scalar field $\varphi$ as
follows,
\begin{equation}\label{Aofpvarphiprofinal}
A(\varphi)=e^{\alpha \varphi}\, ,
\end{equation}
where $\alpha$ is defined as,
\begin{equation}\label{alphaofphi}
\alpha=-2\sqrt{\frac{\pi}{3}}\, ,
\end{equation}
and in the scalar-tensor literature the function $\alpha
(\varphi)$ is,
\begin{equation}\label{alphaofvarphigeneraldef}
\alpha(\varphi)=\frac{d \ln A(\varphi)}{d \varphi}\, ,
\end{equation}
so in our case,
\begin{equation}\label{alphaofphifinalintermsofvarphi}
a(\varphi)=\alpha=-2\sqrt{\frac{\pi}{3}}\, .
\end{equation}
Accordingly, the potential and it's first derivative with respect
to the Einstein frame canonical scalar field $\varphi$ are,
\begin{equation}\label{potentialfinalofvarphi}
V(\varphi)\simeq \frac{\lambda}{\xi^2}\left(e^{2\alpha \varphi}-1
\right)^2\, ,
\end{equation}
and,
\begin{equation}\label{derpotentialfinalofvarphi}
V'(\varphi)\simeq \frac{4 \alpha \lambda}{\xi^2}e^{2\alpha
\varphi}\left(e^{2\alpha \varphi}-1 \right)\, .
\end{equation}
\begin{figure}[h!]
\centering
\includegraphics[width=20pc]{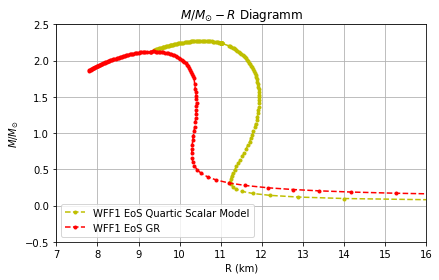}
\includegraphics[width=20pc]{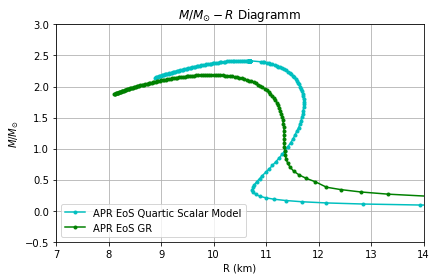}
\includegraphics[width=20pc]{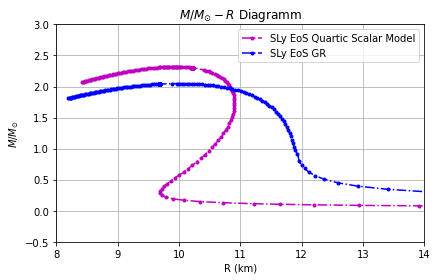}
\caption{$M-R$ graphs for the quartic order scalar model model
results and the GR case for the WFF1 EoS (upper left), the APR EoS
(upper right), and the SLy EoS (bottom plot). The $y$-axis in all
plots corresponds to $M/M_{\odot}$, where $M$ is the Jordan frame
ADM mass, while the $x$-axis corresponds to the Jordan frame
circumferential radius of the NS in kilometers.} \label{plot1}
\end{figure}
The potential (\ref{potentialfinalofvarphi}) is quite well known
in cosmological contexts and yields viable inflationary
phenomenology (see Eq. (43) of Ref. \cite{Mishra:2018dtg}). As it
is known from Ref. \cite{Mishra:2018dtg}, the quartic order scalar
inflationary model yields a viable inflationary phenomenology for,
\begin{equation}\label{inflconstraint}
\frac{\lambda M_p^4}{4\xi^2}\sim 9.6\times 10^{-11}M_p^4\, ,
\end{equation}
where $M_p=\frac{1}{\sqrt{8\pi G}}$. Making the correspondence in
Geometrized units and using the notation of this section, in our
notation the constraint (\ref{inflconstraint}) becomes,
\begin{equation}\label{constraintfinal}
\frac{\lambda}{\xi^2}\sim 16 \pi \times \left(1.51982\times
10^{-13}\right)\, ,
\end{equation}
so by using $\lambda=0.1$ for quartic order scalar model
phenomenological reasoning \cite{Mishra:2018dtg}, this yields $\xi
\sim 11.455\times 10^4$,
 hence, for the NS study we shall use the value $\xi \sim 11.455\times 10^4$, which yields a viable inflationary
phenomenology.

With regard to the values of the coupling parameter $\xi$, there
are some limits coming from different quantum field theories in
curved spacetime \cite{serg2,serg1}. Indeed, we consider quartic
order scalar model scalar sector which basically is just sector of
some realistic grand unified theory (GUT) in curved spacetime.
There are different possibilities for GUTs at the very early
Universe. Let us mention two classes of such GUTS:
asymptotically-free GUTs and finite GUTs, for which some
estimations for value of $\xi$ can be done. For other types of GUT
less strict estimations maybe also developed. Specifically, by
taking renormalization group arguments at high energy, one can
estimate the effective coupling $\xi$ at very high curvature
corresponding to the high energy or very early Universe. It turns
out that this effective coupling constant $\xi$ at high energy
depends on the specific class of GUT under consideration. For
instance, for asymptotically free finite grand unified theories
$\xi$ most usually tends to the value $\xi=1/6$, the conformal
coupling value. In addition, for some asymptotically free GUTs,
$\xi$ can be arbitrarily large and its explicit value depends on
initial conditions \cite{serg2,serg1,serg3}. For such class of
GUTs the large behavior  of $\xi$ is apparent  for $\xi\sim
11.455\times 10^4$, the choice used in this work. In this case,
the constraint (\ref{approxmain2}) is always satisfied for all the
values of the Einstein frame scalar field $\varphi$ (see relation
(\ref{oneplusxiphisquare})), but the first constraint is not
automatically satisfied, namely Eq. (\ref{approxmain1}). Moreover,
for different finite GUTs one has the following possibilities at
high energy: $\xi$ tends to the conformal coupling value $1/6$
(asymptotical conformal invariance), or $\xi$ tends to very large
value, or $\xi$ is arbitrary and it is not defined by quantum
field considerations. Eventually, the last two possibilities for
the values of $\xi$ again support our choice for large $\xi$ under
consideration in our estimation. Note that for effective quantum
field theory we do not have such asymptotic limitations for $\xi$
and it should be derived by cosmological considerations.

It is apparent that for $\xi \sim 11.455\times 10^4$, the
constraint (\ref{approxmain2}) is always satisfied for all the
values of the Einstein frame scalar field $\varphi$ (see relation
(\ref{oneplusxiphisquare})), but the first constraint is not
automatically satisfied, namely Eq. (\ref{approxmain1}). Thus,
when the numerical results are obtained, which will deliver the
values of the scalar field $\varphi$, one must use Eq.
(\ref{oneplusxiphisquare}) to transform the Einstein frame scalar
field to the Jordan frame expression $\xi \phi^2$, and thus
eventually the following constraint must be satisfied for all the
numerical values of the Einstein frame scalar field $\varphi$ that
will be delivered from the numerical code,
\begin{equation}\label{numericalcodeapprox}
\xi^2\phi^2=\xi\,\left(e^{\frac{2\sqrt{16\pi}}{\sqrt{12}}\,
\varphi}-1\right)\gg 1\, .
\end{equation}
Let us now derive the (TOV) equations for a spherically symmetric
compact stellar object, described by the spherically symmetric
static spacetime metric,
\begin{equation}\label{tov1}
ds^2=-e^{\nu(r)}dt^2+\frac{dr^2}{1-\frac{2
m}{r}}+r^2(d\theta^2+\sin^2\theta d\phi^2)\, ,
\end{equation}
where the function $m(r)$ describes the gravitational mass of the
stellar object confined inside a radius $r$.
\begin{figure}[h!]
\centering
\includegraphics[width=20pc]{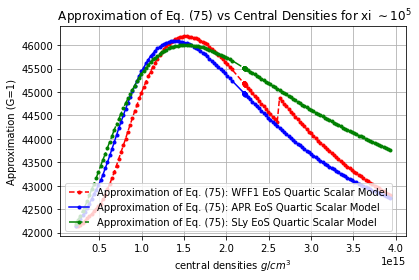}
\caption{The values of the quantity $\xi^2\phi^2$ in the $y$-axis
in Geometrized units, and in the $x$-axis the corresponding
central densities in CGS units, in order to quantitatively verify
the constraint (\ref{numericalcodeapprox}), for $\xi \sim
11.455\times 10^4$.} \label{plot2}
\end{figure}
Assuming the Geometrized physical units system in which $c=G=1$,
the field equations for the Einstein frame action corresponding to
the rescaled canonical scalar field $\varphi$, for the spherically
symmetric metric (\ref{tov1}) read,
\begin{equation}\label{tov2}
\frac{d m}{dr}=4\pi r^2 A^4(\varphi)\epsilon+2\pi
r(r-2m)\omega^2+4\pi r^2V(\varphi)\, ,
\end{equation}
\begin{equation}\label{tov3}
\frac{d\nu}{dr}=4\pi r\omega^2+\frac{2}{r(r-2m)}\Big{[}4\pi
A^4(\varphi)r^3P-4\pi V(\varphi) r^3\Big{]}+\frac{2m}{r(r-2m)}\, ,
\end{equation}
\begin{equation}\label{tov4}
\frac{d\omega}{dr}=\frac{r
A^4(\varphi)}{r-2m}\Big{(}\alpha(\varphi)(\epsilon-3P)+4\pi
r\omega(\epsilon-P)\Big{)}-\frac{2\omega
(r-m)}{r(r-2m)}+\frac{8\pi \omega r^2 V(\varphi)+r\frac{d
V(\varphi)}{d \varphi}}{r-2 m}\, ,
\end{equation}
\begin{equation}\label{tov5}
\frac{dP}{dr}=-(\epsilon+P)\Big{[}\alpha (\varphi)\omega+2\pi r
\omega^2+\frac{m-4\pi r^3(-A^4P+V)}{r(r-2m)}\Big{]}\, ,
\end{equation}
\begin{equation}\label{tov5a}
\frac{d\varphi }{dr}=\omega\, ,
\end{equation}
where the function $\alpha (\varphi)$ is defined in Eq.
(\ref{alphaofphifinalintermsofvarphi}), and the potential and its
derivative are defined in Eqs. (\ref{potentialfinalofvarphi}) and
(\ref{derpotentialfinalofvarphi}) respectively. Also and the
pressure $P$ and energy density $\epsilon$ in the TOV equations
are the Jordan frame quantities. Clearly, the GR limit of the TOV
equations is obtained by setting $A=1$, $\omega=0$, $\alpha
(\varphi)=0$, $V=0$ and of course $\frac{d V}{d \varphi }=0$. The
TOV equations for the scalar-tensor theory in the Einstein frame
must be solved for the following initial conditions,
\begin{equation}\label{tov8}
P(0)=P_c\, ,
\end{equation}
\begin{equation}\label{tov9}
m(0)=0\, ,
\end{equation}
\begin{equation}\label{tov10}
\nu(0)\, ,=-\nu_c
\end{equation}
\begin{equation}\label{tov11}
\varphi(0)=\varphi_c\, ,
\end{equation}
\begin{equation}\label{tov12}
\omega (0)=0\, .
\end{equation}
Near the center the pressure, mass and the metric function have
the following Taylor expansions,
\begin{equation}\label{tov13}
P(r) \simeq P_c -(2\pi)(\epsilon_c+P_c) \left(
P_c+\frac{1}{3}\epsilon_c \right) r^2 + O(r^4)\, ,
\end{equation}
\begin{equation}\label{tov14}
m(r) \simeq \frac{4}{3}\pi\epsilon_cr^3 + O(r^4)\, ,
\end{equation}
\begin{equation}\label{tov15}
\nu(r) \simeq \nu_c + 4\pi \left(P_c+ \frac{1}{3}\epsilon_c
\right)r^2 + O(r^4)\, .
\end{equation}
In the present section we shall perform a numerical integration of
the TOV equations, using the initial conditions above and using a
double shooting method in order to obtain the optimal central
values for the variable $\nu_c$ and the value of the scalar field
$\varphi_c$ at the center of the NS, which make the scalar field
$\varphi(r)$ and the metric function $\nu(r)$ vanish at the
numerical infinity which is chosen to be $r\sim 67.943$ km in the
Einstein frame. The gravitational mass of the star will be assumed
to be the ADM mass measured in solar masses. Let us derive the ADM
mass for the scalar-tensor theory at hand, and specifically the
Jordan frame mass. Let us introduce some notation at this point,
so we define $\mathcal{K}_E$ and $\mathcal{K}_J$ as follows,
\begin{equation}\label{hE}
\mathcal{K}_E=1-\frac{2  m}{r_E}\, ,
\end{equation}
\begin{equation}\label{hE}
\mathcal{K}_J=1-\frac{2 m_J}{r_J}\, ,
\end{equation}
and note that we use Geometrized units. The functions
$\mathcal{K}_E$ and $\mathcal{K}_J$ are related as follows,
\begin{equation}\label{hehjrelation}
\mathcal{K}_J=A^{-2}\mathcal{K}_E\, ,
\end{equation}
while the radii are related as follows,
\begin{equation}\label{radiiconftrans}
r_J=A r_E\, .
\end{equation}
 Also the Jordan frame ADM
mass is,
\begin{equation}\label{jordaframemass1}
M_J=\lim_{r\to \infty}\frac{r_J}{2 }\left(1-\mathcal{K}_J \right)
\, ,
\end{equation}
while the Einstein frame ADM mass is,
\begin{equation}\label{einsteiframemass1}
M_E=\lim_{r\to \infty}\frac{r_E}{2 }\left(1-\mathcal{K}_E \right)
\, .
\end{equation}
Relation (\ref{hehjrelation}) asymptotically yields,
\begin{equation}\label{asymptotich}
\mathcal{K}_J(r_E)=\left(1+\alpha(\varphi(r_E))\frac{d \varphi}{d
r}r_E \right)^2\mathcal{K}_E(r_E)\, ,
\end{equation}
where from now on $r_E$ will denote the Einstein frame radius at
large distances, and $\frac{d\varphi }{dr}=\frac{d\varphi
}{dr}\Big{|}_{r=r_E}$. Hence, by combining the above, after some
simple algebra we obtain the Jordan frame ADM mass in terms of the
Einstein frame ADM mass,
\begin{equation}\label{jordanframeADMmassfinal}
M_J=A(\varphi(r_E))\left(M_E-\frac{r_E^{2}}{2 }\alpha
(\varphi(r_E))\frac{d\varphi
}{dr}\left(2+\alpha(\varphi(r_E))r_E\frac{d \varphi}{dr}
\right)\left(1-\frac{2 M_E}{r_E} \right) \right)\, ,
\end{equation}
where $\frac{d\varphi }{dr}=\frac{d\varphi }{dr}\Big{|}_{r=r_E}$
For the numerical calculation we shall first find the Einstein
frame ADM mass, and by using Eq. (\ref{jordanframeADMmassfinal})
we shall calculate the Jordan frame ADM mass, expressed in solar
masses, in Geometrized units. With regard to the Einstein frame
radius of the neutron star $R_s$, will be obtained from the
numerical code, which is the value of the radius at the point of
the star where the pressure becomes zero $P(R_s)=0$. Thus the
value of the radius at the surface, namely $R_s$, will be
transformed to the Jordan frame value  in kilometers, using the
relation,
\begin{equation}\label{radiussurface}
R=A(\varphi(R_s))\, R_s\, ,
\end{equation}
and expressing eventually $R$ in kilometers, where $\varphi(R_s)$
is the value of the scalar field at the surface of the neutron
star. In addition, for the delivered values of the Einstein frame
scalar field $\varphi$, we must validate that the constraint of
Eq. (\ref{numericalcodeapprox}) is indeed satisfied by our
numerical results (keeping Geometrized units for this task for
convenience). The numerical code is a hybrid version of the freely
available code pyTOV-STT \cite{niksterg} which is a python 3 based
double shooting numerical code appropriately modified to
accommodate the scalar potential. The TOV equations shall be
integrated for both the interior and the exterior of the star
(where $\epsilon=P=0$ and only the potential survives), using the
``LSODA'' numerical method, and for the exterior integration
caution is needed because the numerical infinity must be
appropriately chosen.

For the choice $r\sim 67.94378528694695$km, the numerical results
are optimal. We shall use a piecewise polytropic EoS with the
initial pressure $p_1$ and the parameters $\Gamma_2$, and
$\Gamma_3$ corresponding to the values of three distinct EoSs, the
WFF1 \cite{Wiringa:1988tp} which is a variational method EoS, the
SLy \cite{Douchin:2001sv} which is a potential method EoS, and the
APR EoS \cite{Akmal:1998cf}. Also, with regard to the quartic
order scalar model potential parameters $\lambda$ and $\xi$, we
shall fix $\lambda=0.1$ for quartic order scalar model
phenomenological reasoning \cite{Mishra:2018dtg}, and $\xi$ will
be chosen $\xi \sim 11.455\times 10^4$, since this value is the
most relevant to the inflationary theory, due to the fact that for
this value the inflationary theory becomes compatible with the
latest Planck data on inflation \cite{Akrami:2018odb}. In the
following we shall present the results of our analysis, which will
basically consist of the $M-R$ graphs for all the studied cases, a
direct comparison to the $M-R$ graphs of GR for all the
aforementioned EoSs. Also we shall explicitly check whether the
Jordan frame constraint (\ref{numericalcodeapprox}) holds true for
all the obtained numerical results.

Let us now present the results of our numerical analysis and we
also discuss the features of NSs when these are described by the
quartic order scalar model. Let us start with the $M-R$ graphs,
and in Figs. \ref{plot1} we present the $M-R$ graphs for all the
EoSs and we directly compare the quartic order scalar model
results with the GR results, for $\xi \sim 11.455\times 10^4$. In
the upper left we present the GR and quartic order scalar model
$M-R$ graph for the WFF1 EoS, with the GR curve being red while
the quartic order scalar model curve being yellow. In the upper
right plot of Fig. \ref{plot1} we present the APR $M-R$ graph for
both GR (green curve) and the quartic order scalar model (cyan),
while in the bottom plot of Fig. \ref{plot1} we present the $M-R$
graph for the SLy EoS for GR (blue) and the quartic order scalar
model (purple).

Clearly, the $M-R$ curves describe quite well the $M\sim 1.6
M_{\odot}$ area, where the radius is expected to be constrained in
the range $R=10.68^{+15}_{-0.04}$km  \cite{Bauswein:2017vtn}. More
importantly, the WFF1 EoS in the context of GR, which was excluded
by the GW170817 constraints \cite{Bauswein:2017vtn}, in the case
of the quartic order scalar model NS model, becomes compatible
with the GW170817 constraints. In all the plots, the $y$-axis
corresponds to $M/M_{\odot}$, where $M$ is the ADM mass, while the
$x$-axis corresponds to the Jordan frame circumferential radius of
the NS in kilometers. In Table \ref{table1} we collect all the
maximum masses and the corresponding radii, for all the EoSs. The
resulting picture is interesting since the only fundamental scalar
experimentally verified in nature, the quartic order scalar model,
leads to maximum NS radii which are higher but quite close to the
corresponding GR limit, for all the EoSs studied in this paper.
More importantly, for all the maximum radii static NS
configurations appearing in Table \ref{table1}, the corresponding
radii are also compatible with the GW170817 constraint, which
dictates that the radii corresponding to the maximum mass of a
static NS must be larger than $R=9.6^{+0.14}_{-0.03}$km.
\begin{table}[h!]
  \begin{center}
    \caption{\emph{\textbf{Maximum Masses and the Corresponding  of Static NS for the Quartic Order Scalar Model and for GR}}}
    \label{table1}
    \begin{tabular}{|r|r|r|r|}
     \hline
      \textbf{Model}   & \textbf{APR EoS} & \textbf{SLy EoS} & \textbf{WFF1 EoS}
      \\  \hline
      \textbf{GR} & $M_{max}= 2.18739372\, M_{\odot}$ & $M_{max}= 2.04785291\, M_{\odot}$ & $M_{max}= 2.12603999\, M_{\odot}$
      \\  \hline
      \textbf{Quartic Scalar $\xi\sim 10^5$} & $M_{max}= 2.41722734\,M_{\odot}$ & $M_{max}= 2.32004294\,M_{\odot}$
      &$M_{max}= 2.27222814\, M_{\odot}$ \\  \hline
       \textbf{Corresponding Quartic Scalar Model Radii} & $R= 10.35829055$km &
$R= 9.91149993$km
      &$R= 10.74919782$km \\  \hline

    \end{tabular}
  \end{center}
\end{table}
Furthermore, in Fig. \ref{plot2} we present the values of the
quantity $\xi^2\phi^2$ in the $y$-axis in Geometrized units, and
in the $x$-axis the corresponding central densities in CGS units,
in order to quantitatively verify the constraint
(\ref{numericalcodeapprox}), for $\xi \sim 11.455\times 10^4$.
Obviously, the constraint (\ref{numericalcodeapprox}) is well
satisfied by our numerical results, for $\xi\sim 10^4$. In
addition, we stress here that the constraint (\ref{approxmain2})
is automatically satisfied for the values of $\xi$ we used in this
article.

\section*{Concluding Remarks}

In this paper we studied the effects of a quartic order scalar
model in static NSs, in the Einstein frame. We investigated
quantitatively the properties of NSs when the quartic order scalar
model potential is present. By using the approximation
$\xi^2\phi^2\gg 1$ in the Jordan frame, which holds true for large
$\xi$ values, we derived the Einstein frame TOV equations
corresponding to a static spherically symmetric spacetime. With
regard to the EoS, we used piecewise polytropic EoSs, with the
central part being described by the SLy, APR and WFF1 EoS, which
are well known successful EoSs. We numerically integrated the TOV
equations using a double-shooting ''LSODA'' python 3 based
numerical code, and we extracted the NS masses and radii, as well
as the values of the scalar field in the Einstein frame. With
regard to the mass, we used the ADM mass in the Jordan frame,
which is particularly useful for static spacetimes, and with
regard to the radii, we converted the Einstein frame radii to
their Jordan frame counterparts. We numerically solved the TOV
equations for both the interior and the exterior of the NS, and
the numerical study of the exterior was particularly demanding,
needing a careful choice of the numerical infinity in order to
optimize the central values of the scalar field and of the metric
function that yield asymptotically at the numerical infinity, zero
values for the scalar field and the metric function. The whole
numerical study was performed for a large number of central
densities. Using the masses and radii data, we constructed the
$M-R$ graphs for the quartic order scalar model. For all the
studied EoSs, the $M-R$ graphs were compatible with the
observational constraints imposed by the GW170817 event which
require the radius of a static $M\sim 1.6 M_{\odot}$ neutron star
to be larger than $R=10.68^{+15}_{-0.04}$km
\cite{Bauswein:2017vtn} and in addition the radius of a static
neutron star corresponding to the maximum mass of the star to be
larger than $R=9.6^{+0.14}_{-0.03}$km. Moreover, an important
outcome of this work is the fact that although the WFF1 EoS  was
excluded for static neutron stars in the context of GR, for the
quartic order scalar model neutron star model it provides
realistic static NS configurations, which are compatible with both
the aforementioned constraints of the GW170817 event
\cite{Bauswein:2017vtn}. We also carefully examined whether the
initial approximation we made in the Jordan frame, namely
$\xi^2\phi^2\gg 1$ holds true, and as we showed the approximation
holds true. What now remains is to investigate how large are the
baryon masses for such quartic order scalar model neutron stars,
in order to understand the mass limit for which neutron stars will
collapse to black holes. In the same line of research, it is
important to calculate the causal mass limit for the present
theory, with variable speed of sound, and investigate the
qualitative behavior of the model. These two tasks may provide
useful insights towards understanding how large can a neutron star
mass be, and when do neutron stars can collapse to black holes.
The latter can serve perhaps for a lower limit of astrophysical
black holes masses, and may offer new insights to the difficult
question, how small can astrophysical black holes masses be and
how large a neutron star mass can be. Finally, the present work
indicates that the quartic order scalar model predicts maximum
radii for NS that are off the mass-gap region, thus it would be
interesting to investigate whether string-corrected scalar field
models, such as the Einstein-Gauss-Bonnet models
\cite{Hwang:2005hb,Kanti:2015pda,DeLaurentis:2015fea,Yi:2018dhl,Kleihaus:2019rbg,Bakopoulos:2019tvc},
may provide maximum masses inside the mass-gap region. We hope to
address this issue in future works.

\section*{Acknowledgments}

V.K. Oikonomou is indebted to N. Stergioulas and his MSc student
Vaggelis Smyrniotis for the many hours spend on neutron star
physics discussions and for sharing his professional knowledge on
numerical integration of neutron stars in python. This work was
supported by MINECO (Spain), project PID2019-104397GB-I00.

\end{document}